\documentclass{elsart}
\usepackage{graphicx}
\usepackage{amssymb}
\usepackage{natbib}
\begin{document}
\begin{frontmatter}
\title{Noise sensitivity of portfolio selection under various risk measures}
\author[addr1,addr2]{Imre Kondor\corauthref{cor}},
\author[addr3,addr2]{Szil\'ard Pafka},
\author[addr4]{G\'abor Nagy}
\address[addr1]{Collegium Budapest -- Institute for Advanced Study, \\
Szenth\'aroms\'ag u. 2., H--1014 Budapest, Hungary}
\address[addr2]{Department of Physics of Complex Systems, E\"otv\"os University,\\
P\'azm\'any P.\ s\'et\'any 1/a, H-1117 Budapest, Hungary} 
\address[addr3]{Paycom.net\\
2644 30th Street, 2nd Floor, Santa Monica, CA 90405, USA}
\address[addr4]{Risk Management Department, CIB Bank,\\
Medve u.\ 4--14., H--1027 Budapest, Hungary}
\corauth[cor]{Corresponding author. E-mail: kondor@colbud.hu.
Phone: +36-1-224-8313. Fax: +36-1-375-9539.}
\begin{abstract}
We study the sensitivity to estimation error of portfolios optimized under various risk measures, including variance, absolute deviation, expected shortfall and maximal loss. We introduce a measure of portfolio sensitivity and test the various risk measures by considering simulated portfolios of varying sizes $N$ and for different lengths $T$ of the time series. We find that the effect of noise is very strong in all the investigated cases, asymptotically it only depends on the ratio $N/T$, and diverges at a critical value of $N/T$, that depends on the risk measure in question. This divergence is the manifestation of a phase transition, analogous to the algorithmic phase transitions recently discovered in a number of hard computational problems. The transition is accompanied by a number of critical phenomena, including the divergent sample to sample fluctuations of portfolio weights. While the optimization under variance and mean absolute deviation is always feasible below the critical value of $N/T$, expected shortfall and maximal loss display a probabilistic feasibility problem, in that they can become unbounded from below already for small values of the ratio $N/T$, and then no solution exists to the optimization problem under these risk measures. Although powerful filtering techniques exist for the mitigation of the above instability in the case of variance, our findings point to the necessity of developing similar filtering procedures adapted to the other risk measures where they are much less developed or nonexistent. Another important message of this study is that the requirement of robustness (noise-tolerance) should be given special attention when considering the theoretical and practical criteria to be imposed on a risk measure. 
\end{abstract}
\end{frontmatter}


\section{Introduction}
Risk is one of the central concepts in finance playing a prominent role in investment decisions, asset allocation, risk management, and regulation alike. Despite its fundamental importance, no universally accepted measure exists for its quantitative characterization. Risk measures used by practitioners, implemented in risk management software packages, embodied in regulation, or applied in theoretical considerations range from rules of thumb and ad hoc recipes to standard statistical measures and sophisticated axiomatic constructs. A comprehensive treatment of risk measures should embed them in the wider context of economic theory. This would entail, among other things, the clarification of their relationship to utility functions, and a discussion of how the choice of one or another risk measure reflects a set of implicit assumptions about the nature of the underlying processes, investors and markets. Our goals are much more modest here. Taking a pragmatic approach, we will content ourselves with considering a few concrete risk measures which are common in practice and the academic literature, and will focus on a sole issue related to them, namely their sensitivity to estimation error (noise). 

Most practitioners tend to regard risk as a given number, or perhaps a set of a few numbers. This is an oversimplified view, however. Ultimately, risk figures result from evaluating some estimators whose input data come from empirical observations on the market. As the observed samples (or time series) are always finite, sample to sample fluctuations are inevitable. The risk characteristics of any financial asset or portfolio are, therefore, never single, well defined numbers, but random variables. Ideally, their probability densities would be so sharply peaked that even the observation of a single sample (a finite length time series) would give a fair representation of the whole distribution. The fact is, however, that the size of typical banking portfolios and the lengths of available time series are such that this is almost never the case, and the problem of estimation error can almost never be neglected in a finance context. The fundamental problem we face here is one of information deficit. In typical situations the amount of data needed for a faithful reconstruction of the underlying stochastic process far exceeds the amount of data available. Under these conditions, our stochastic models will inevitably contain some amount of measurement error and the risk estimates will be noisy. 

The effect of noise is much more serious when we want to {\it choose} a portfolio that is optimal under certain criteria than when we merely assess the risk in an {\it existing} portfolio. The same risk measure may perform fairly satisfactorily as a {\it diagnostic} tool and fail miserably as a {\it decision making} tool. The focus of this paper will be on the decision making side: we will study the noise sensitivity of risk measures in {\it portfolio selection}, as opposed to risk assessment.

The problem of estimation error is, of course, not new. It dates back to the very beginnings of Markowitz' rational portfolio selection theory \citep{MarkowitzPap,MarkowitzBook}, and over the decades a huge number of papers have been devoted to its various aspects. Most of the literature is concerned with portfolios composed of assets that obey normal or Gaussian statistics, and discusses the noise sensitivity of the natural risk measure associated with Gaussian portfolios, namely, the variance
\citep[e.g.][]{MVErrFrankf,MVErrDick,MVErrJobson,MVErrElton,MVErrEun,MVErrChan}.  

In addition to studying the effect of noise, a number of filtering schemes have also been devised to remove at least a significant part of the estimation noise from portfolio selection. These filtering procedures include single- and multi-factor models, \citep[for a review, see e.g.][]{EltonGruberBook}, Bayesian estimators \citep[e.g.][]{BayJorion,BayFrost,BayLedoit}, and more recently, tools taken over from random matrix theory \citep{RMTBouchaud2}. 

Despite the fact that portfolio optimization based on several alternative risk measures has been introduced in the literature
\citep{MadKonno,MinimaxYoung,ESUryasev,SpecAcerbiBookChapt}
and become utilized in practice \citep{MadAlgoRes,MadAlgoOptim},
the literature on the effect of noise on risk measures other than the variance is limited. (This is certainly not true of the widespread risk measure value at risk or VaR. VaR, as a quantile is, however, not a convex measure, therefore no meaningful comparison can be made between the noise sensitivity of VaR and that of the convex measures treated here. The case of VaR will not be considered in this paper.)

The purpose of this paper is to systematically compare the noise sensitivity of the following risk measures: standard deviation, absolute deviation, expected shortfall, and its limiting case, maximal loss. For this purpose we test them under idealized, "laboratory" conditions: 

\begin{itemize}
\item[--] in order to have a complete control over the stochastic process and avoid other sources of noise (e.g. non-stationarity), we use simulated data;
\item[--] for the sake of simplicity, we assume that the distribution of return fluctuations is normal;
\item[--] in order to get rid of the even harder problem of expected returns, we concentrate on the minimal risk portfolio;
\item[--] and finally, we allow unlimited short selling. 
\end{itemize}

Under these simplifying assumptions we find that the effect of noise is so strong in all the investigated cases that it actually diverges at a critical value of the ratio of the portfolio size $N$ and the length $T$ of the time series. This divergence is, in fact, the manifestation of an algorithmic phase transition, and constitutes a new type of these transitions that have recently been discovered in a number of hard optimization and decision problems. \citep[For an excellent review, see][]{MezardMontanari}. The recognition of the divergence of estimation error and the identification of this divergence as an algorithmic phase transition are the central results of this paper.  With these we have established a link between the problem of portfolio selection and an important new development in computer science as well as a highly developed chapter of statistical physics, the theory of phase transitions. This provides access to a plethora of powerful concepts and methods, such as scaling ideas, critical phenomena, critical exponents, universality, finite size scaling, etc., and may eventually suggest new filtering and optimization strategies.  

The divergent estimation error in the whole portfolio is accompanied by even stronger sample to sample fluctuations of the portfolio weights. In the course of this study we have also encountered a surprising feasibility problem: While the optimization under variance and mean absolute deviation is always feasible below the critical value of $N/T$, expected shortfall and maximal loss may become unbounded from below in some samples already for small values of the ratio $N/T$, and then no solution exists to the optimization problem under these risk measures. 

A remark on the role of the simplifying assumptions above is in order. As the instability of portfolio selection is an information-deficit catastrophe, we do not believe that the use of real market data, or non-stationary time series, or fat-tailed distributions, or the introduction of a risk free asset and a constraint on expected return would qualitatively modify our conclusions. This is not at all true of the last assumption, the lack of a constraint on short selling. It is evident that a ban on short selling (or any other set of constraints that would render the domain over which we seek an optimum finite) would automatically eliminate the possibility of a divergence. It would then seem that our results refer to a completely unrealistic case. We insist, however, that it is useful to consider this unrealistic case first, because it helps understand the root of the instability and identify the strong residual fluctuations that reflect this instability even after the constraints are reintroduced. This is analogous to the theory of phase transitions where it proved to be essential to consider the limit of infinite volume, even though no real physical system has infinite volume. In the case of portfolio selection, when the ratio $N/T$ is not far enough from its critical value, the residual sample to sample fluctuations of the weights are so strong even in a finite volume that they make the task of rational portfolio selection largely illusory. 

The plan of the paper is as follows. In Section \ref{chapvariance} we look into the noise sensitivity of a classical risk measure, the variance. We introduce the mean relative estimation error (the sample average of the ratio of the noisy standard deviation and the true one) as a measure of the noise-induced sub-optimality of a portfolio and display an exact analytic expression for this quantity that shows that the effect of noise diverges as we approach the limit $N/T=1$ from below. In addition to the estimation error, we also study the instability of portfolio weights.

Section \ref{chapotherriskmeas} is devoted to the study of the noise sensitivity of mean absolute deviation (MAD), expected shortfall (ES), and maximal loss (ML). They are all found to display similar, but even stronger sensitivity to estimation error than the variance. In addition, we notice that optimization under ES and ML is not always feasible even for small values of  $N/T$, we derive a closed analytic formula for the probability of the existence of a solution for ML, and determine this probability for ES via numerical simulations.

The paper ends on a short Summary.

The exposition will be informal throughout. No lengthy mathematical derivations will be presented, and the results will be illustrated or supported by simulations.

\section{Noise sensitivity of variance}
\label{chapvariance}

\subsection{The Markowitz problem}

The fluctuations of financial returns form a multivariate stochastic process. The simplest model for their pdf is provided by a multivariate normal distribution. This picture goes back to \citet{Bachelier}, and it has remained the standard textbook model of financial markets even to this day. Real-life portfolios conform to this model to various degrees, depending on the assets, liquidity, the time horizon, etc.  For the sake of simplicity, we assume that our portfolio is multivariate normal.

The problem of rational portfolio selection was formulated by \citet{MarkowitzPap} as a tradeoff between reward and risk.  Reward is usually measured in terms of return or log return. For a Gaussian portfolio variance is an essentially unique measure of risk; any other reasonable measure of risk is necessarily proportional to it. Rational investors want to minimize their risk given a certain fixed expected return. In mathematical terms the task consists in finding the minimum of the quadratic form
\begin{equation}
\label{eqsigmaportf}
\sigma_P^2=\sum_{i,j=1}^N w_i \sigma_{ij} w_j,
\end{equation}
over the weights $w_i$, given the constraints
\begin{equation}
\label{eqcondbudget}
\sum_{i=1}^N  w_i =1
\end{equation}
and 
\begin{equation}
\label{eqcondreturns}
\sum_{i=1}^N  w_i \mu_i=\mu,
\end{equation}
where $\sigma_P$ is the standard deviation of the portfolio, $\sigma_{ij}$ the covariance matrix, $w_i$ the weight of asset $i$ in the portfolio ($i=1...N$), $\mu$ the expected return on the portfolio (given), and $\mu_i$ the expected return on asset $i$.  We assume that there is absolutely no constraint on short selling, so the weights $w_i$ can be of either sign and of any absolute value. This is, of course, quite unrealistic, partly for legal, partly for liquidity reasons, but this idealized setup is where we can display the instability of portfolio selection in its purest form.

The classical optimization problem above can be solved analytically \citep{MertonEffFront}. (However, if short selling is excluded or other linear constraints are introduced, it becomes a quadratic programming problem.)

Expected returns are notoriously hard to determine on short time horizons with any degree of reliability. As our objective in this paper is to study the noise sensitivity of risk measures, we want to simplify the task and omit the constraint on return. That is, we wish to focus on the minimal risk portfolio. At first sight this may seem rather pointless. We note however, that there are special tasks (benchmarking, index tracking) where this is precisely what one wishes to do.
The solution is then:
\begin{equation}
w_i^{*}=\frac{\sum_{j=1}^N\sigma_{ij}^{-1}}{\sum_{j,k=1}^N\sigma_{jk}^{-1}}\,.
\end{equation}

It is important to note that the optimal weights are given here in terms of the inverse covariance matrix. Since the covariance matrix has, as a rule, a number of small eigenvalues, any measurement error will get amplified and the resulting portfolio will be sensitive to noise. This is the fundamental reason for the difference between portfolio selection and risk assessment of a given portfolio; in the latter case, the covariance matrix does not need to be inverted.

\subsection{Empirical covariance matrices}

The covariance matrix has to be determined from measurements on the market. From the returns  $x_{it}$ observed at time $t$ we get the maximum likelihood estimator:
\begin{equation}
\label{eqsigmaempirical}
\sigma_{ij}={\frac{1}{T}}\sum_{t=1}^{T} x_{it}x_{jt}
\end{equation}
(assuming that the expected values are known to be zero). 

For a portfolio of $N$ assets the covariance matrix has $O(N^2)$ elements. The time series of length $T$ for $N$ assets contain $NT$ data. In order for the measurement to be precise, we need $ N^2\ll NT$, that is $N\ll T$. Bank portfolios may contain hundreds or thousands of assets. The length of available time series depends on the context, but it is always bounded. If we talk about a stock portfolio for example, it is hardly reasonable to go beyond four years, that is $T\sim$ 1000 (some of the stocks may not have been present earlier, economic or regulatory environment may change, etc.). Therefore, $N/T \ll 1$ rarely holds in practice. As a result, there will be a lot of noise in the estimate. Considering analogous situations in the theory of phase transitions, we expect that for large enough $N$ and $T$ the error depends only on the ratio $N/T$, which we describe by saying that it scales in $N/T$.

The problem we have just described is one of several manifestations of the ''curse of dimensionality''. Economists have been struggling with this problem for many years
\citep{EltonGruberBook}. Since the root of the problem is lack of sufficient information, the remedy is to inject external information into the estimate. This means imposing some structure on the matrix $\sigma$, which introduces bias, but the beneficial effect of noise reduction may compensate for this. These filtering procedures, Bayesian estimators, etc. have a large body of literature of their own.  As our primary concern here is the comparison of the noise sensitivity of risk measures, we do not enter into a discussion of the procedures by which this sensitivity is reduced. 

\subsection{A measure of the effect of noise on the optimal portfolio}

For the purposes of quantitative analysis we need to introduce a measure that characterizes the effect of noise or estimation error on portfolio selection. One could use various metrics defined over the space of covariance matrices for this purpose: one may define a distance between matrices, or between the spectra of these matrices, etc. We believe, however, that the most relevant measure is the relative sub-optimality, or relative risk increment of the portfolio $q_0-1$, where
\begin{equation}
\label{eqq0def}
q_0^2=\frac{\sum_{ij} w^{*}_i\sigma^{(0)}_{ij}w^{*}_j}
{\sum_{ij} w^{(0)*}_i\sigma^{(0)}_{ij}w^{(0)*}_j}.
\end{equation}

Here $\sigma^{(0)}$ is the ''true'' covariance matrix (the empirical one will be called $\sigma$), and $w^{(0)*}$ and $w^{*}$ are the weights of the portfolios optimized under $\sigma^{(0)}$, and $\sigma$, respectively. The square root of the denominator in Eq.\ (\ref{eqq0def}) is the true risk (standard deviation) of the portfolio, while the square root of the numerator is the risk we run when using the weights derived from the empirical covariance matrix.

This measure was introduced in \citet{mynoisy1}.  It assumes, of course, that we know the true process. By a slight extension of the definition, one can make it applicable also in the context of empirical data \citep{mynoisy2}, but this will not concern us in this paper.

Eq. (\ref{eqq0def}) implicitly refers to a given sample, a segment of length $T$ from the time series. Therefore, $q_0$ itself is a random variable. In this paper we will mainly be concerned with its average over the samples $\bar{q}_0$ and will only perform a preliminary study of its full distribution.

\subsection{Divergent estimation error under variance}

\subsubsection{The simplest covariance matrix}

Let us imagine we have a portfolio of standard, independent, normal variables. The corresponding covariance matrix is the simplest concievable: just the unit matrix. If we now generate $N$ series of length $T$ of these variables and determine their covariance matrix through the formula Eq.\ (\ref{eqsigmaempirical}), we will, however, not recover the unit matrix unless we let $T\to\infty$ for $N$ fixed. Instead, we find a much more complicated structure that will fluctuate from sample to sample. Accordingly, the average estimation error $\bar{q}_0$ for these "empirical" covariance matrices will also fluctuate and will always be larger than unity for any finite $T$. In the limit of large $N$ and $T$ values, however, such that their ratio is fixed, $N/T < 1$, methods borrowed from the theory of random matrices allow us to exactly determine $\bar{q}_0$ and obtain the strikingly simple result:

\begin{equation}
\label{q0anal}
\bar{q}_0=\frac{\textstyle 1}{\sqrt{1-\frac{N}{T}}}.
\end{equation}

This formula dates back to a discussion between two of the present authors (I.\ K. and S.\ P.) and G.\ Papp and M.\ Nowak, and was published in \citet{mynoisy2}, \citet{RMTNowak} and \citet{RMTPappGvel}.

It is well known that the rank of the empirical covariance matrix is the smaller of $N$ and $T$. When $T$ becomes smaller than $N$, the covariance matrix develops zero eigenvalues, and the portfolio optimization problem becomes meaningless. Eq.\ (\ref{q0anal}) shows that as we approach the limit $N/T=1$, the mean relative error in the portfolio diverges. 

\begin{figure}
\begin{center}
\includegraphics[scale=0.55]{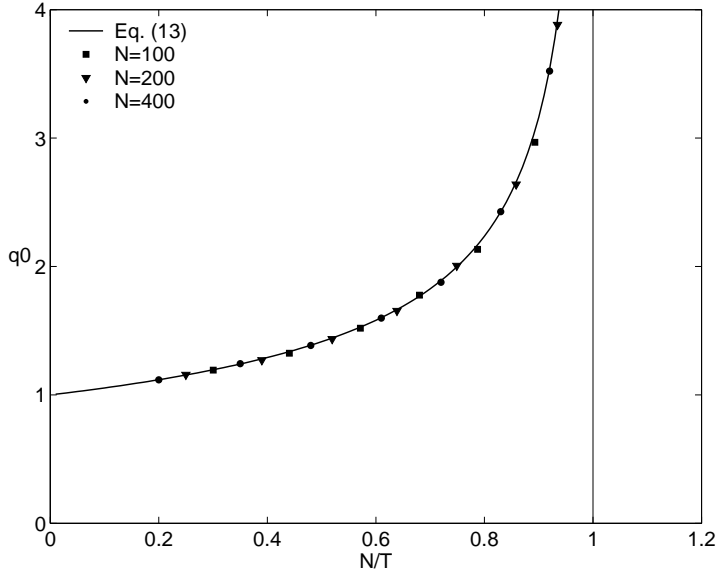}
\end{center}
\caption{Mean $q_0$ as a function of $N/T$ for different values of $N$ (simulation results). The data points collapse on the curve given by Eq.\ (\ref{q0anal}) (also shown in the figure).
\label{figq0TN}}
\end{figure}

Eq.\ (\ref{q0anal}) is an asymptotic result, valid only in the limit $N,T\to\infty$. The convergence is, however, very fast. In Fig.\ \ref{figq0TN} we show the results of simulations for the average of $q_0$ for various $N$ and $T$ values. The data points nicely fit the theoretical result (corresponding to the limit $N,T \to\infty$ ) already for moderate values of these parameters, and demonstrate the scaling in the variable $N/T$.

\begin{figure}[h!]
\begin{center}
\includegraphics[scale=0.55]{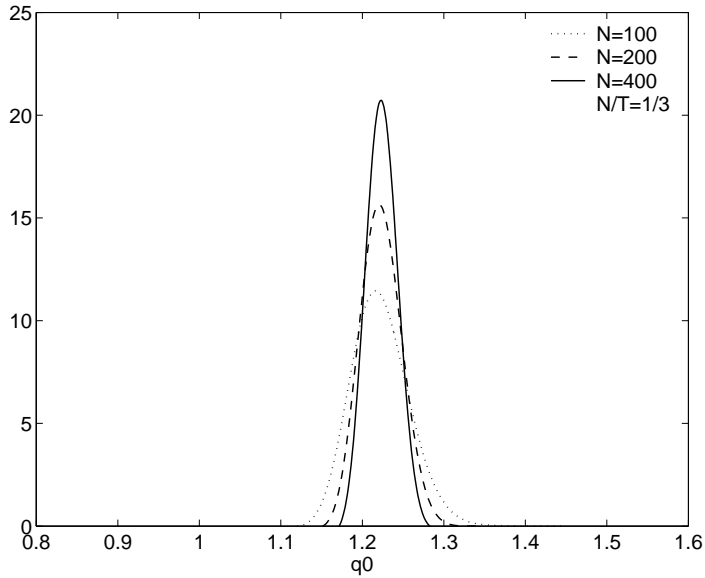}
\end{center}
\caption{Distribution of $q_0$ for increasing $N$ with $N/T$ fixed
(simulation results).
\label{figq0hist}}
\end{figure}

In contrast to the average of $q_0$, no analytic result is available for the distribution of the estimation error, so here we have to resort to numerical simulations. These show that the distribution of $q_0$ becomes sharper and sharper with increasing $N$, if we stay at a fixed distance from the critical point, i.e. keep $N/T$ fixed. This is illustrated in Fig.\ \ref{figq0hist} where the histogram of $q_0$ is displayed for three different values of $N$, with $N/T=1/3$ fixed. It can be seen that the maximum of the distribution is located above 1.2, so the typical relative error in the optimal portfolio is over 20\% 
even for this relatively small value of the ratio $N/T$. It is also seen that although the distribution becomes sharper as $N$ increases, it remains fairly wide even for the largest value of $N=400$ . This means that not only the typical values of the error are high, but this error also wildly fluctuates from sample to sample. 

Numerical experiments also demonstrate that in the opposite limit, i.e. for a fixed portfolio size $N$ and for the ratio $N/T$ going to unity from below, the distribution of $q_0$ becomes wider and wider. In particular, the standard deviation of $q_0$ shows an even stronger divergence than its average. A detailed study of the distribution of the relative error demands a numerical effort which goes beyond the scope of this paper and will be the subject of a subsequent publication.

The above findings clearly suggest the picture of a phase transition. (For a first orientation on phase transitions see http://en.wikipedia.org/wiki/Phase\_transition, or \citet{ StanleyPhaseTrans}.) As we approach the critical value 1 of the control parameter $N/T$, various quantities (e.g. the average estimation error, but also all the higher moments of its distribution) display critical behaviour: they diverge as a power of the distance from the critical point. Admittedly, this phase transition is of a somewhat peculiar type, in that the divergent fluctuations remain infinitely large even as we go into the phase on the other side of the critical point, $N/T >1$. This is due to the simple quadratic objective function and should not concern us here. We are not going to develop this phase transition analogy in further detail here, but will use it as a heuristic guide in the following. 

One may wonder to what extent the result for the divergent estimation error depends on the extremely simple covariance model we chose at the beginning of this subsection. The phase transition analogy suggests that its essential features do not depend on it. The relevant principle we have to quote here is universality. It is a general feature of critical phenomena that the exponents of the power laws describing the divergence of various quantities around a critical point do not depend on the fine details of the models in question, a feature ultimately going back to the central limit theorem. Therefore we expect that although the prefactor of the power law in Eq.\ (\ref{q0anal}) may change as we go from the trivial covariance matrix considered so far to more elaborate market models, the exponent of the divergence will remain the same. We do not know the precise boundaries of the universality class within which this critical index remains constant, but numerical experiments on a number of simple models (including the \citet{FactorNoh} model that reproduces the known spectral properties of empirical covariance matrices reasonably well) have convinced us that this universality class may well extend to models that display some of the characteristic features observed on real markets.The study of the domain of universality will be published elsewhere.

\subsection{Fluctuating weights}

We have seen that the effect of noise on the average relative standard deviation can be very strong, especially for large portfolios and not sufficiently long time series. The average $q_0$ is, however, only a global measure of the effect of noise. A more detailed characterization can be obtained by considering the optimal weights belonging to the empirical covariance matrix. As this matrix fluctuates from sample to sample, so do the weights. 

\begin{figure}
\begin{center}
\includegraphics[scale=0.55]{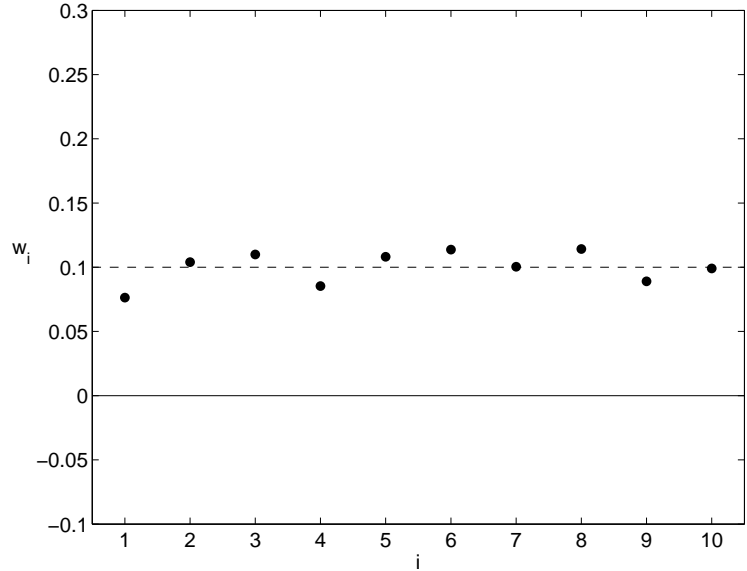}
\end{center}
\caption{Optimal portfolio weights $w_i^{*}$ ($i=1...N$) obtained from an estimated  correlation matrix compared to the \ ''true'' weights (dashed line) for $N=10$, $T=500$.
\label{figwiNsmall}}
\end{figure}

\begin{figure}
\begin{center}
\includegraphics[scale=0.55]{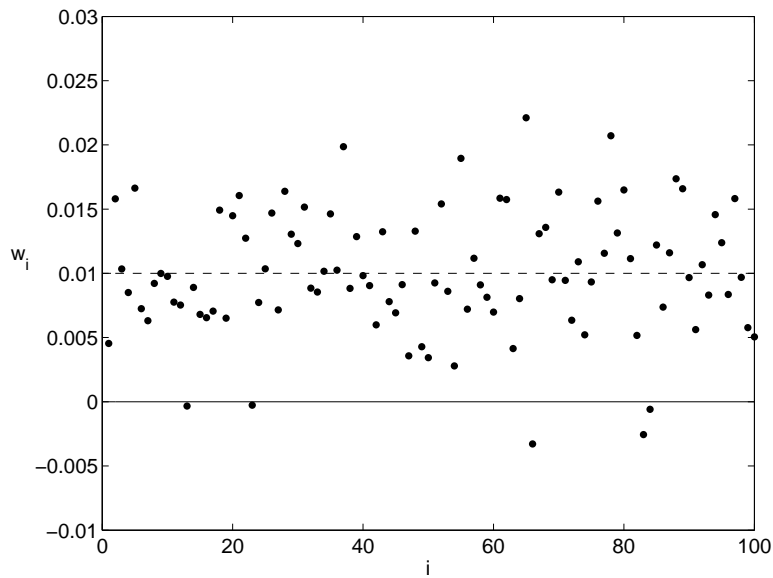}
\end{center}
\caption{Optimal portfolio weights $w_i^{*}$ ($i=1...N$) obtained from an estimated correlation matrix compared to the ''true'' weights (dashed line) for $N=100$, $T=500$.
\label{figwiNlarge}}
\end{figure}

In Fig.\ \ref{figwiNsmall} we exhibit the optimal empirical weights for a small portfolio ($N=10$) of independent standard Gaussian items, while in Fig.\ \ref{figwiNlarge} we show the same for a larger portfolio ($N=100$). The simulation results displayed have been obtained from a randomly chosen sample of length $T=500$, so the ratio $N/T$ is quite far from the critical threshold $N/T=1$ even for the larger portfolio. As in this illustrative example all the assets are assumed to be completely equivalent, the ''true'' weights are all equal ($1/N$). They are also shown in the figures for comparison. The deviation of the empirical weights from their true value is striking.

\begin{figure}
\begin{center}
\includegraphics[scale=0.55]{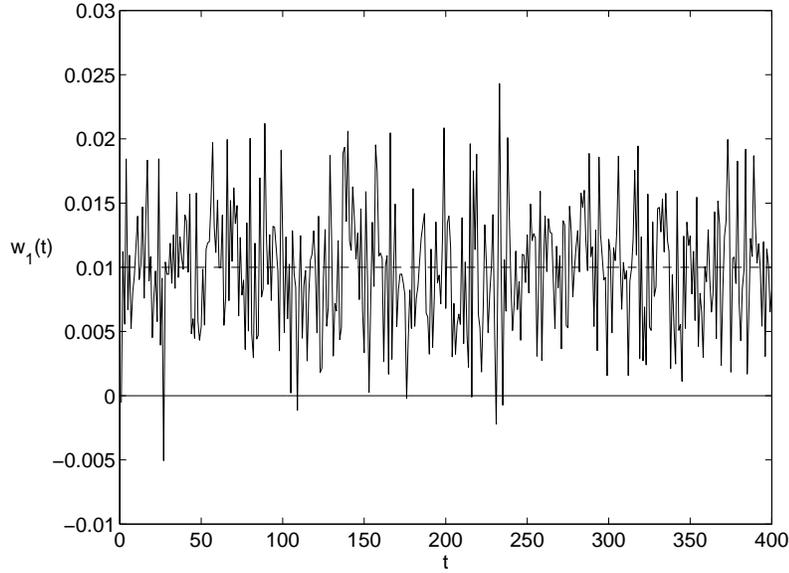}
\end{center}
\caption{$w_1^{*}$ as a function of time for $N=100$, $T=500$, when time windows are non-overlapping (the dashed line shows the ''true'' value).
\label{figw1new}}
\end{figure}

\begin{figure}
\begin{center}
\includegraphics[scale=0.55]{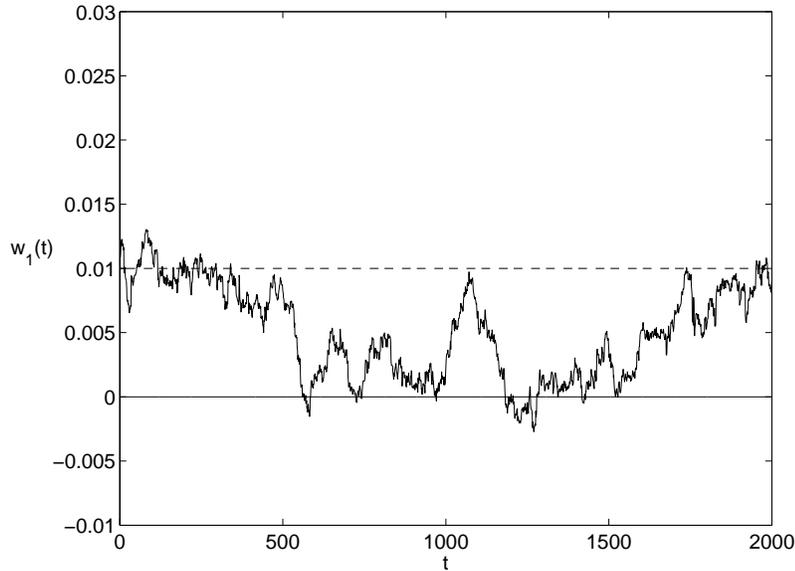}
\end{center}
\caption{$w_1^{*}$ as a function of time for $N=100$, $T=500$, when the time window steps forward one unit at a time (the dashed line shows the ''true'' value).
\label{figw1shift1}}
\end{figure}

Figs.\ \ref{figw1new} and \ref{figw1shift1} demonstrate the fluctuations of the empirical weights in time for $N=100$ and $T=500$. In Fig.\ \ref{figw1new} we show the path of a given weight calculated from non-overlapping time windows of length $T$, while in Fig.\ \ref{figw1shift1} we display the same for a window of the same length $T$, but stepping forward one unit at a time. In the former case the weight undergoes wild fluctuations, in the latter its steps are strongly correlated, hence the path is much smoother, but it stays mostly far from its true position. Evidently, neither of these is very promising from a risk management point of view. In the former case, the hypothetical portfolio should be totally reorganized every period $T$, in the latter it would seem deceptively stable from one day to the next, but it would be far from its true optimal composition most of the time.  

The distribution of weights shows a divergence similar to (and even stronger than) the distribution of $q_0$ .  The detailed study of this distribution would lead us far from the main line of this paper, so we content ourselves with a few remarks. As the wild fluctuations of the weights set in already quite far from the critical region (already for small values of the ratio $N/T$), it is hard to see how this optimization problem could offer any basis for rational decision making. The truth is that in its present form it does not. In real life, however, we never have unlimited short positions, so the domain over which the optimum is sought is bounded. This not only prevents $q_0$ from blowing up, but also tames the fluctuations of the weights considerably. Instead of running away to large negative values, a number of the weights will then stick to the boundaries of the allowed region. The spontaneous reduction of the portfolio size is well known in the finance literature (see e.g. \citet{NuoptBook} p.\ 8 and subsequent pages). In the context of the instability described above, we see that the appearance of a large number of weights sitting on the boundaries is a remnant of the critical behaviour that is now smoothed out by the finite volume. This will not, however, lend too much credibility to the optimized portfolio: the set of weights that stick to the walls will strongly fluctuate from sample to sample. We see then that the ban on short selling, or any other constraints that render the domain of the problem finite, will mask rather than resolve the instability.

The real remedy is, of course, filtering. A number of these noise reduction techniques are available in the literature \citep{EltonGruberBook} and widely used in practice. The discussion of filtering is not a subject of this paper, however. In the next section we wish to consider the noise sensitivity of some alternative risk measures for which hardly any filtering procedures exist.  For a fair comparison between the noise sensitivities of various risk measures, we had, therefore, to consider unfiltered results for the variance.

\section{Alternative risk measures}
\label{chapotherriskmeas}

For a normal distribution the typical intensity of fluctuations can be completely characterized by a single number, the variance or the standard deviation. For other distributions the variance will not suffice. For long-tailed distributions that asymptotically fall off like a power-law, variance can be particularly misleading as a risk measure. Real-life portfolios often display this long-tailed behavior, and consequently should not be optimized under variance. Alternative risk measures abound both in practice and the literature. We cannot cover all of them here, but focus on a few that have some practical and/or theoretical importance.

\subsection{Absolute deviation} 

Mean absolute deviation (MAD, the expected value $\mathrm{E}(|x|)$ of the absolute value of fluctuations) is an obvious alternative to standard deviation \citep{KonnoL1}.  Some risk management methodologies \citep[e.g.][]{MadAlgoOptim} do actually use MAD to characterize the fluctuation of portfolios, which offers a huge computational advantage in that the resulting portfolio optimization task can be solved by linear programming. However, the effect of estimation noise on MAD has been largely ignored in the literature \citep[except][]{MadEsterror}. Preliminary results of our study of the noise sensitivity of MAD have appeared in \citet{KondorPafkaAD}.

Given a time series of finite lenght $T$, the objective function to minimize is
\citep{MadKonno}
\begin{equation}
\frac{1}{T} \sum_t \Big|\sum_i w_i x_{it}\Big|,
\end{equation}
subject to the usual budget constraint $\sum_i w_i=1$.
(We again disregard the constraint on expected return.)

This is equivalent to the following linear programming problem:
\begin{eqnarray}
\min && \frac{1}{T} \sum_t u_t \\
\mathrm{s.t. \;} && u_t+ \sum_i w_i x_{it} \ge  0 \label{ADLPut1} \\
&& u_t -\sum_i w_i x_{it} \ge 0 \label{ADLPut2} \\
&& \sum_i w_i = 1 
\end{eqnarray}
where the minimization is carried out over $w_i$ and the additional variables $u_t$, $t=1...T$, while Eqs.\ (\ref{ADLPut1}--\ref{ADLPut2}) represent $2T$ constraints, a pair for each $t=1...T$.

Now we apply the same simulation-based strategy as with the variance. We generate artificial data of a known structure, which we choose here, for the sake of simplicity, to be independent standard normal again. Since for normal fluctuations 
$\mathrm{E}(|x|)=\sqrt{\frac{2}{\pi}}\,\sigma$, the ratio $q_0$ of the MAD of the portfolio constructed by the above optimization procedure ($w^{*}_i $) and the MAD of the ''true'' optimal portfolio ($w^{(0)*}_i$) is equal to the ratio of the standard deviations of these portfolios:
\begin{equation}
q_{0,MAD}^2=\frac{\sum_{i} {w^{*}_i}^2}
{\sum_{i} {w^{(0)*}_i}^2}.
\end{equation}
By symmetry, the ''true'' optimal weights are all equal ($w^{(0)*}_i =\frac{1}{N}$), the ratio $q_0$ which can be used to characterize the sub-optimality of the portfolio obtained from time series of finite length is then
\begin{equation}
q_{0,MAD}^2=N{\sum_{i} {w^{*}_i}^2}.
\end{equation}

\begin{figure}
\begin{center}
\includegraphics[scale=0.55]{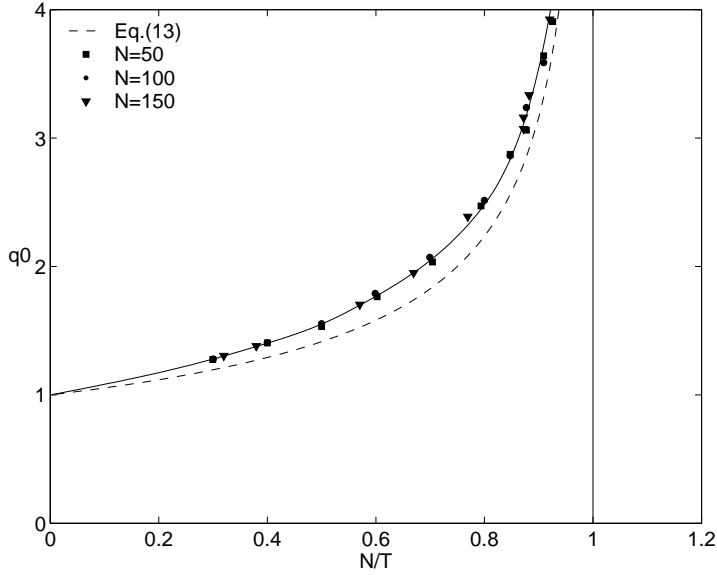}
\end{center}
\caption{The sample average of $q_{0,MAD}$ as a function of $N/T$ (simulation results). The data points collapse on a curve situated above the curve obtained previously for the variance, also shown in the figure (with dashed line).
\label{figq0AD}}
\end{figure}

We have performed simulations for this quantity for various $N/T$ values and show the results in Fig.\ \ref{figq0AD}. It is clear that the data points collapse on a single curve again which shows that (the mean of) $q_{0,MAD}$ scales in $N/T$ in this case, too. Also shown in the figure are results obtained for the variance. As Fig.\ \ref{figq0AD} clearly demonstrates, $\bar{q}_{0,MAD}$ lies above the corresponding curve for the variance, which means that MAD as a risk measure is more sensitive to noise than the variance. (This result is consistent with \citet{MadEsterror}.) However, both curves diverge at the same point $N/T=1$, and asymptotically close to this critical point they behave the same way, so the critical exponent of  $\bar{q}_{0,MAD}$ is the same for the variance and MAD.

A geometric interpretation of the enhanced sensitivity of MAD compared to the variance has been given in \citet{KondorPafkaAD}.  The essential point is that the iso-risk surfaces of MAD are polyhedra, as opposed to the ellipsoidal level surfaces of the variance. The solution of the optimization problem is found at the point where this risk-polyhedron first touches the plane of the budget constraint. This happens typically at one of the corners of the risk-polyhedron. If we construct this polyhedron from finite length time series, it will inevitably contain some estimation error or noise. Accordingly, the shape and/or the position of the risk-polyhedron will change from sample to sample. As a result, the solution will {\it jump} to a new corner of the new polyhedron. This discontinuity is the basic reason for the enhanced sensitivity. 

For  fixed $N$ and $T$ going to infinity, the polyhedron goes over into the ellipsoid of constant variance, and the difference between MAD and variance-optimized portfolios disappears. For finite $T$, however, the piecewise linear character persists, and it is precisely this, otherwise attractive, feature of MAD that makes it prone to estimation noise.

\subsection{Expected shortfall}

Expected shortfall (ES) \citep[see e.g.][]{SpecAcerbiBookChapt} is the mean loss beyond a high threshold $\beta$ {\it defined in probability} (not in money). For continuous pdf's it is the same as the conditional expectation beyond the value at risk (VaR) quantile, therefore sometimes it is also called Conditional VaR or CVaR, but for discrete distributions (such as the histograms in empirical studies) CVaR and ES are different, see \citet{ESAcerbiCoher} for a careful discussion of the subtle difference between the two. The increasing popularity of ES is due to the fact that it is a coherent (hence also convex) measure \citep{ESAcerbiCoher}, perhaps the simplest and most intuitively appealing of all the coherent measures. (In fact, it also belongs to the special subset of coherent measures called spectral measures \citep{SpecAcerbiBookChapt}.) In addition, \citet{ESUryasev} have shown that the optimization of ES can be reduced to linear programming for which extremely fast algorithms exist. 

The ES objective function (to be minimized over $v$ and the weights $w_i$) is 
\citep{ESUryasev}:
\begin{equation}
\left(v+\frac{1}{(1-\beta)T}\sum_t [-v-\sum_i w_i x_{it}]^{+}\right),
\end{equation}
subject to $\sum_i w_i=1$. (The constraint on expected return will be omitted, as all through this paper.) We have used the notation $[a]^+ = a$, if $a>0$, zero otherwise, and
$\beta$ is the ES threshold.

The optimization task above is equivalent to the following linear programming problem:
\begin{eqnarray}
\min && \left(v+\frac{1}{(1-\beta)T}\sum_t u_t\right) \\
\mathrm{s.t. \;} &&  u_t\ge -v-\sum_i w_i x_{it} \\
&& u_t\ge 0 \\
&& \sum_i w_i = 1 
\end{eqnarray}
where now the minimization is carried out over $w_i$, $v$ and $u_t$, $t=1...T$.

To test the noise sensitivity of ES we use a portfolio of independent standard normal random variables, and, in complete analogy with the previous cases, measure the sub-optimality due to finite $T$ samples in terms of the ratio $q_{0,ES}$ between the risk (as measured by ES) evaluated for the optimal weights obtained for a given sample and the same with uniform weights. 

On the basis of the experience with the previous two risk measures we are prepared to meet difficulties beyond a certain critical value of the ratio $N/T$, but the simulations of $q_{0,ES}$ confront us with a completely unexpected phenomenon: the ES optimization problem above does not always have a solution even for small values of $N/T$!  More precisely, the existence of a solution depends on the sample and thus becomes a probabilistic issue for any $N/T$. The probability of the existence of a solution depends on the parameters ($N$, $T$, and $\beta$) of the problem.  Before setting out to measure the noise sensitivity of ES, we have to clarify this feasibility problem, and map out those regions of parameter space where we can hope to find solutions with a nonnegligible probability. It turns out that the problem is numerically quite demanding, therefore, in order to gain a preliminary orientation, first we consider a special case, when the threshold $\beta$ is very close to one. The point is that if $\beta$ is so close to unity that $(1-\beta)T \le 1$, then only the single worst loss will contribute to ES. This limiting case represents a coherent (spectral) measure in its own right; we will call it maximal loss (ML). The feasibility problem of ES is present in ML too, but it is analytically tractable, so ML provides a convenient laboratory for understanding the phenomenon. This is what we turn to now.

\subsection{Maximal loss}

As a risk measure, maximal loss may appear to be over-pessimistic: we consider the worst loss ever incurred on a portfolio of a given composition, then minimize this loss over the weights. The objective function to be minimized is then
\begin{equation}
\label{Eqminmax}
\max_t \Big(-\sum_i w_i x_{it}\Big),
\end{equation}
subject to the budget constraint $\sum_i w_i=1$.

The minimax problem \citep{MinimaxYoung} so defined is equivalent to the following linear programming task:
\begin{eqnarray}
\min && u \\
\mathrm{s.t. \;} && u \ge -\sum_i w_i x_{it} \\
&& \sum_i w_i = 1.
\end{eqnarray}

The sub-optimality of the portfolio $q_{0,ML}$ is defined as the ratio of the risk evaluated for the optimal weights obtained for a given sample and the same with uniform weights. As we know that the optimization problem is not always feasible, the definition of  $q_{0,ML}$ must be understood conditional on the existence of a solution.

\begin{figure}
\begin{center}
\includegraphics[scale=0.55]{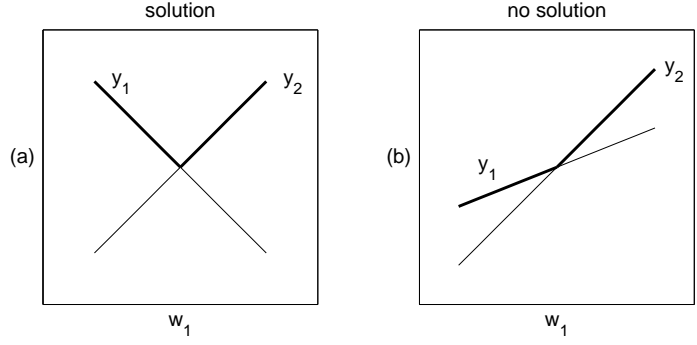}
\end{center}
\caption{$y_1(w_1)$ and $y_2(w_1)$ in the two cases discussed in the text. (a): the ML measure is bounded from below, and a solution exists, (b): the measure is not bounded, there is no solution.
\label{figsurfML}}
\end{figure}

The feasibility problem is rather more transparent in this simplified setting. Consider the trivial case $N=2$, $T=2$. Then we have two linear functions of the weights in the argument of the minimax problem (Eq.\ \ref{Eqminmax}):
\begin{eqnarray}
y_1&=& -w_1 x_{11} - (1-w_1)x_{21}=(x_{21}-x_{11}) w_1 - x_{21} \\
y_2&=& -w_1 x_{12} - (1-w_1) x_{22}=(x_{22}-x_{12}) w_1 - x_{22},
\end{eqnarray}
where we have already eliminated $w_2$ through the budget constraint.
If the slopes of these two straight lines are of opposite sign, i.e.\ if either $x_{21}>x_{11}$ and $x_{22}<x_{12}$, or $x_{21}<x_{11}$ and $x_{22}>x_{12}$, there will be a solution to the minimax problem, otherwise there will not. Fig.\ \ref{figsurfML} shows the arrangement in the two cases. It is easy to see that if the returns $x_{it}$ are independent standard normal random variables, the probability of there being a solution is $1/2$. 

Now we can understand the root of the problem: In contrast to the variance and absolute deviation which are positive semidefinite by definition, the ML measure is not always bounded from below, and when it is not, no solution will be found. It is also clear that a ban on short selling, i.e. constraining all the weights to be non-negative, will automatically eliminate the problem.

For generic $N,T$ we will have $T$ random planes in an $N$-dimensional space, and the condition for the existence of a solution is that these planes form a convex polytope. Induction on $N$ and $T$ leads to the following formula for this probability:
\begin{equation}
p=\frac{1}{2^{(T-1)}} \sum_{k=N-1}^{T-1} {{T-1}\choose{k}}.
\end{equation}

Although we have been arguing in terms of independent normal variables, a quick symmetry argument shows that the statement is, in fact, valid also for correlated normal, moreover, for elliptical distributions. Indeed, boundedness of the measure cannot depend on the orientation of the coordinate axes, nor on the stretching or shrinking transformations that bring a normal distribution over into an elliptical one.

Problems of an identical nature appeared earlier in the context of stochastic programming and random geometry \citep[see][]{MinimaxTodd,MinimaxProof}.

\begin{figure}
\begin{center}
\includegraphics[scale=0.55]{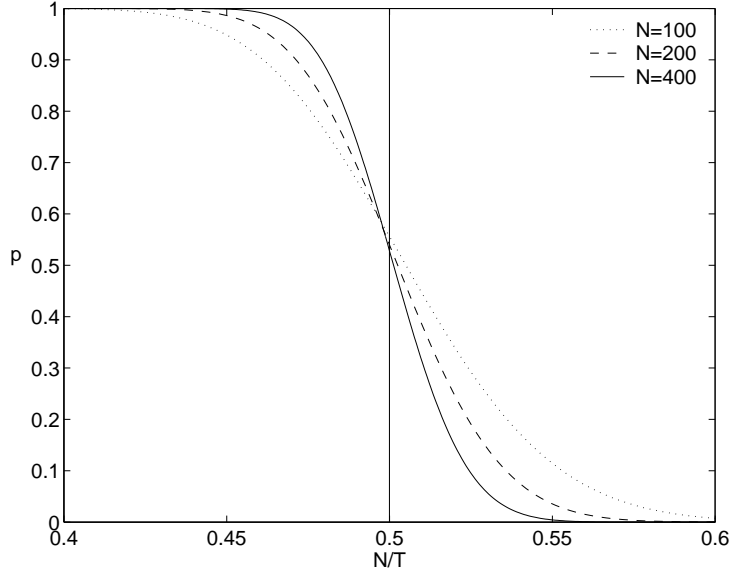}
\end{center}
\caption{Probability $p$ that the minimax problem has solution as function of
$N/T$ for various values of $N$. As $N\to\infty$, $p$ goes over into a step function.
\label{figpML}}
\end{figure}

Fig.\ \ref{figpML} is an illustration of the behaviour of $p$ as function of $N/T$ for various values of $N$. With $N$ increasing, the transition becomes sharper and sharper, until in the limit $N,T\to\infty$ the transition becomes abrupt, and $p$ becomes a step function, as in the case of variance.

For large values of $N$ and $T$ the probability distribution above can be well approximated by the error function:
\begin{eqnarray}
&& p\approx1-\Phi(z), \\
&& \Phi(z)=\int_{-\infty}^{z} \frac{1}{\sqrt{2\pi}}\,e^{-\frac{1}{2}s^2 }\textrm{d}s, \\
&&z=2\left(\frac{N}{T}-\frac{1}{2}\right)\sqrt{T}.
\end{eqnarray}

\begin{figure}
\begin{center}
\includegraphics[scale=0.55]{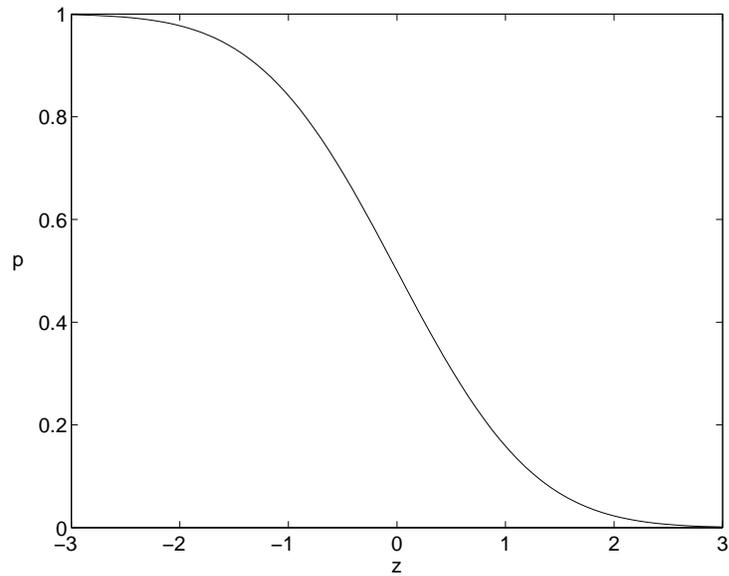}
\end{center}
\caption{Probability $p$ that the minimax problem has solution as function of
$z=2\left(\frac{N}{T}-\frac{1}{2}\right)\sqrt{T}$.
\label{figpMLscaling}}
\end{figure}

In Fig.\ \ref{figpMLscaling} the probability of a solution is plotted against the variable $z=2\left(\frac{N}{T}-\frac{1}{2}\right)\sqrt{T}$. 
When both $N$ and $T$ are large, and $N/T<1/2$, $z$ is large and negative, and the probability of a solution is very close to 1. When $N/T>1/2$, $z$ is large positive and  $p$ is close to zero. The transition is sharper and sharper as $T$ increases. The special value $N/T=1/2$  plays a role similar to the threshold $N/T=1$ we encountered earlier in the case of variance and MAD:  1/2 is the critical point for the ML optimization algorithm.

A comparison between the case of variance and ML optimization may be of interest. 
As the rank of the covariance matrix is the smaller of $N$ and $T$, optimization under variance is always feasible for $N/T <1$, and always meaningless for $N/T>1$. We might have introduced a probability for a solution also for the optimization under variance: this probability would then have a discontinuity at $N/T=1$. The difference in the case of ML is that for finite $N,T$ the transition is continuous, which means that we can never be absolutely sure of the existence of a solution unless $T$ goes to infinity. The convergence is, however, very fast at both ends, so now we know what to expect when we start a simulation for ML at given parameter values.

\begin{figure}
\begin{center}
\includegraphics[scale=0.55]{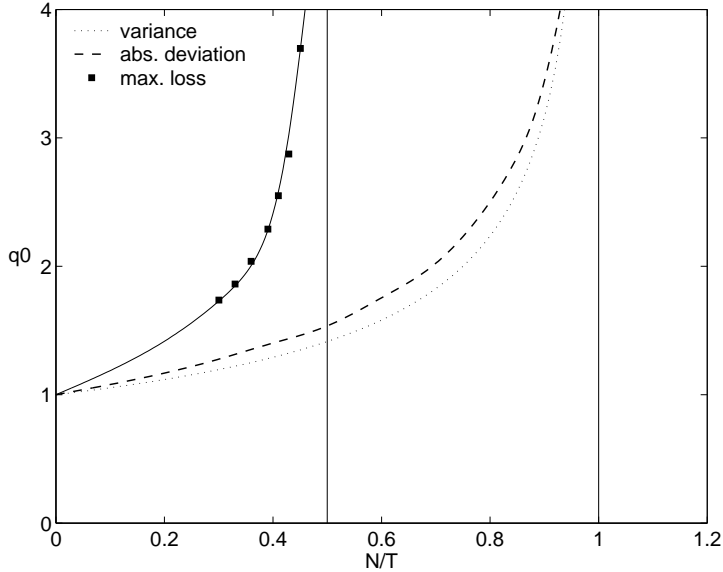}
\end{center}
\caption{$q_{0,ML}$ as a function of $N/T$ (simulation results). The curves
previously obtained for variance and MAD are also shown in the figure for comparison. 
\label{figq0ML}}
\end{figure} 

Assuming that we can find a sufficient number of ''good'' samples for which a solution exists, we can run the simulations and measure $q_{0,ML}$ as a function of $N/T$. We are not surprised to find that $q_{0,ML}$ is much larger than either $q_0$ for the variance or $q_{0,MAD}$  for MAD. Indeed, the iso-risk surfaces of maximal loss are again polyhedra, just like those of MAD, which makes the solution jump about under noise. More importantly, however, the very construction of ML implies that we throw away almost all the data, except the worst ones. No wonder the estimates will be very unstable under these circumstances.

Fig.\ \ref{figq0ML} displays a comparison between the noise sensitivities of variance, MAD and ML. It is clear that $q_{0,ML}$ is the most sensitive of the three, moreover it blows up already at the critical value $N/T=1/2$.

\subsection{The feasibility problem for Expected Shortfall}

\begin{figure}
\begin{center}
\includegraphics[scale=0.55]{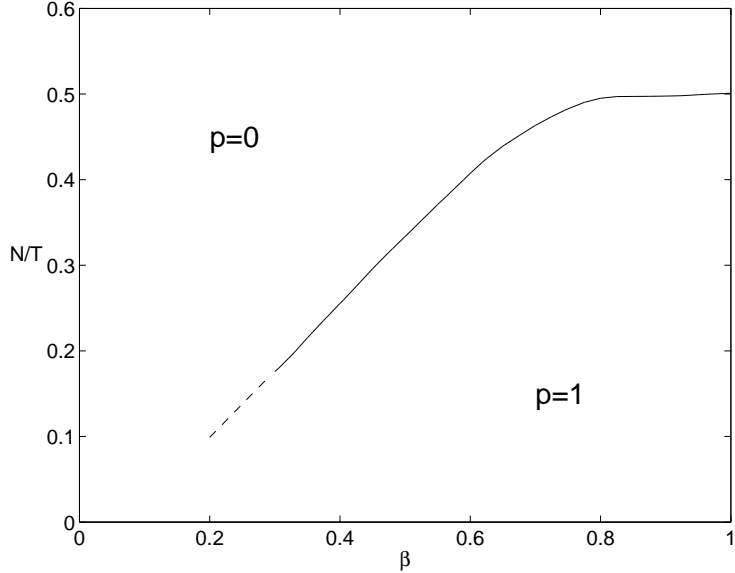}
\end{center}
\caption{Phase diagram showing the $p=0$ and $p=1$ ''phases'' and the transition line between them in the $\beta$--$N/T$ plane ($N,T\to\infty$).
\label{figpESphase}}
\end{figure}

Now we can chart the feasibility map of ES. Extensive numerical experiments suggest the ''phase diagram'' shown in Fig.\ \ref{figpESphase}:  this is the line that separates the region where a solution exists from that where it does not. This feasibility map corresponds to the case  $N,T\to\infty$ with their ratio varying along the vertical axis.  The threshold $\beta$ changes along the horizontal axis. We know from the previous subsection that in the limiting case $\beta=1$ (which is ML) the probability of the existence of a solution drops from one to zero at $N/T=1/2$ as we move upwards along the $\beta=1$ line. As we move away from $\beta=1$, the critical value of $N/T$, where the probability of having a solution has this discontinuity, decreases. This means that for a fixed $N$ we need longer and longer time series to have the same chance of finding a solution. This may be interpreted as ES becoming, in a sense, more sensitive to noise as $\beta$ decreases. Since decreasing $\beta$ corresponds to taking into account more and more data, this result is quite unexpected. 

The precise shape of the phase boundary is difficult to determine, especially for small values of the ratio $N/T$, where the simulations slow down tremendously. For these reasons we have not been able to follow the decline of the phase boundary beyond $\beta$ around 0.3 (althought it is easy to see that the boundary must terminate in the origin, since for $\beta>0$ and $N/T\to 0$ one has ''full information'', therefore $p=1$, while for $N/T>0$ and $\beta\to 0$ the objective function to be minimized becomes linear, thus always unbounded in the weights, therefore $p=0$). Small $\beta$ values, of course, do not have any practical significance, so we feel the essential point about the behaviour of the phase boundary is its downward bend when $\beta$ decreases from around 1.

\begin{figure}
\begin{center}
\includegraphics[scale=0.55]{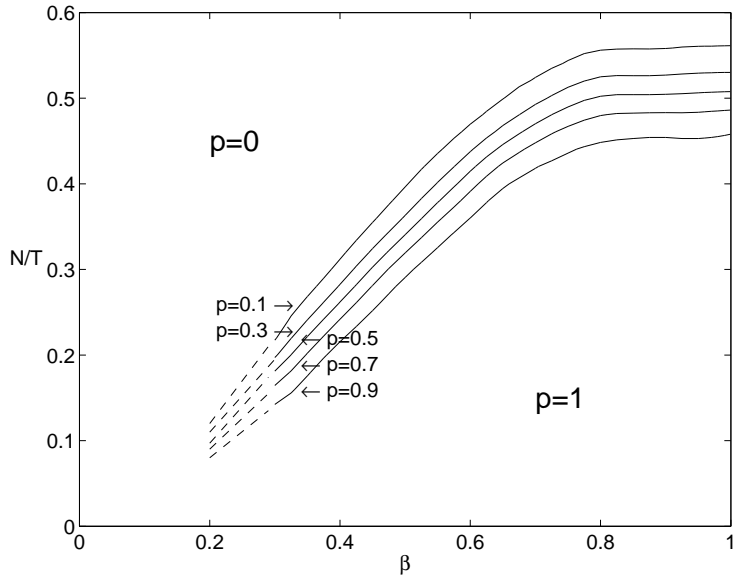}
\end{center}
\caption{Contour lines of $p$ as a function of $\beta$ and $N/T$ for finite $N=100$.
\label{figpESphasefinite}}
\end{figure}

\begin{figure}
\begin{center}
\includegraphics[scale=0.55]{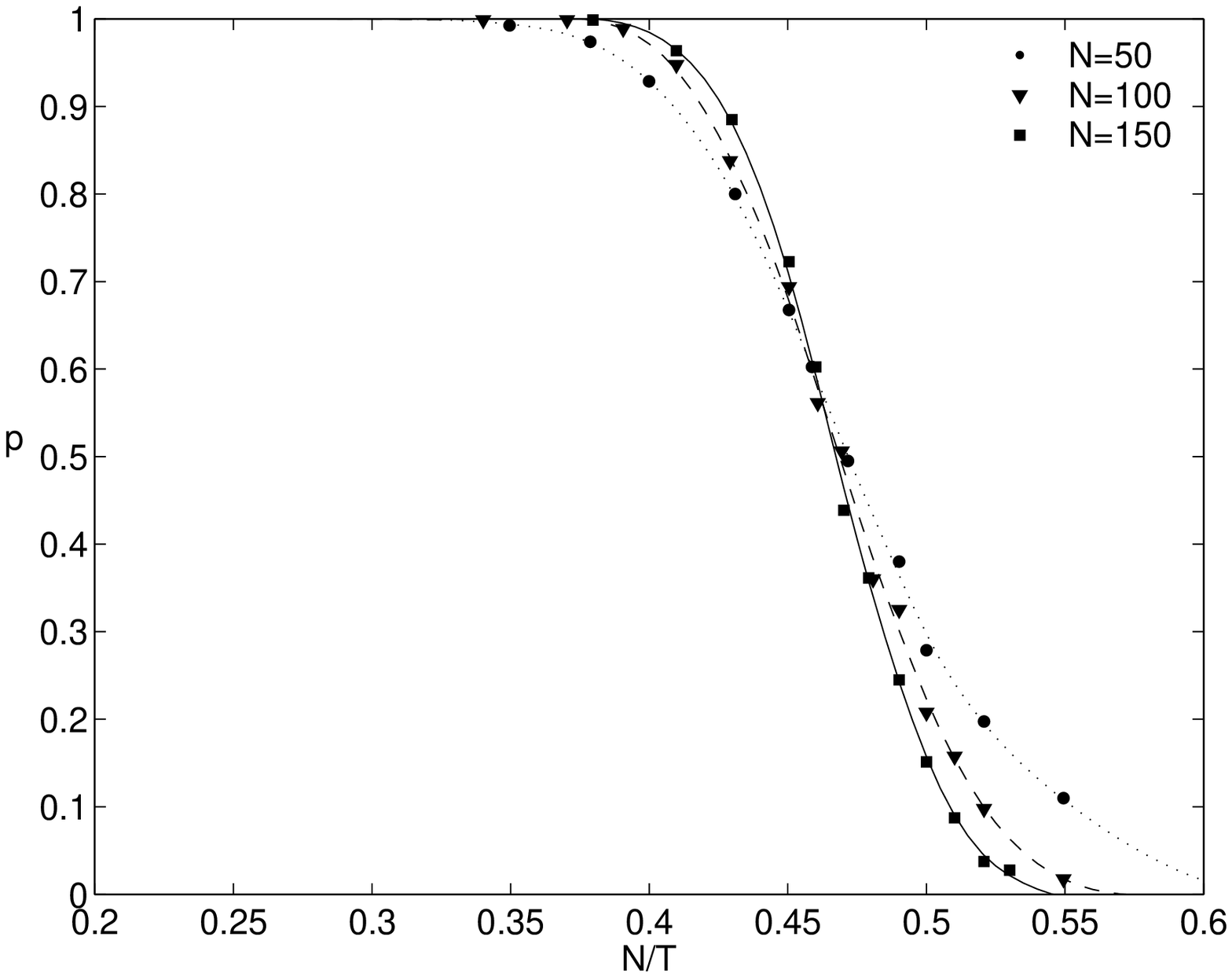}
\end{center}
\caption{Probability $p$ that the optimization of ES is feasible
as a function of $N/T$ for $\beta=0.7$ (for various values of $N$).
As $N\to\infty$, $p$ goes over into a step function.
\label{figpES}}
\end{figure}

For finite $N,T$ values we find that the sharp drop observed in the limit $N,T\to\infty$ goes over into a continuous transition, similar to the case in ML. This is illustrated in Fig.\ \ref{figpESphasefinite}, where the iso\-$p$ lines of the $p(\beta,N/T)$ function are shown for $N=100$, and in Fig.\ \ref{figpES}, where the variation of $p$ with $N/T$ is represented for $\beta=0.7$ fixed (for various values of $N$). Again, the feasibility problem disappears if we ban short selling, just as with ML.

\begin{figure}
\begin{center}
\includegraphics[scale=0.55]{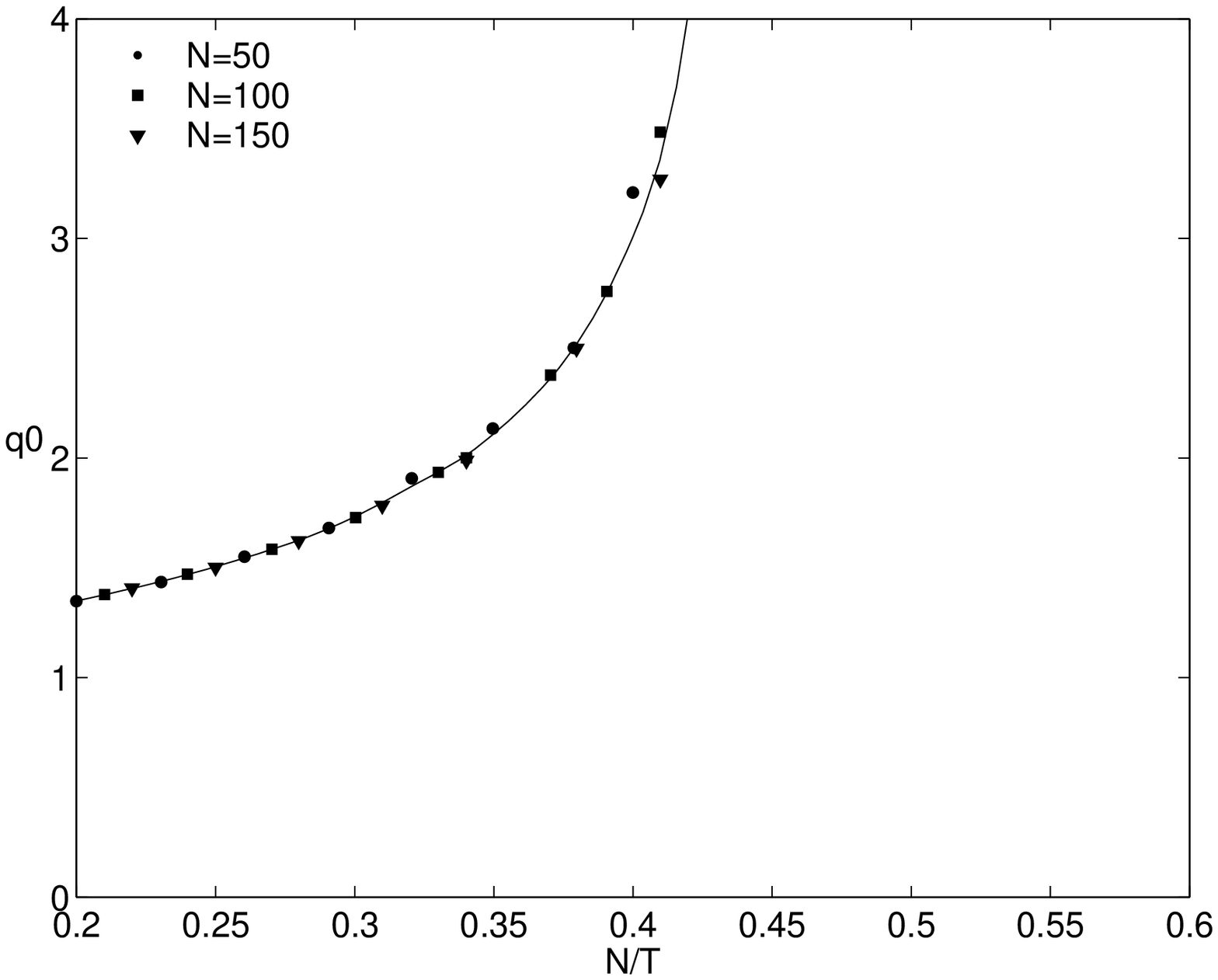}
\end{center}
\caption{$q_{0,ES}$ as a function of $N/T$ for fixed $\beta=0.7$ (for various values of $N$).
\label{figq0ES1}}
\end{figure}

We can also measure the average relative increase of risk due to noise, $\bar{q}_0$. Along a fixed $\beta$ line (assuming a solution exists) as $N/T$ increases, $\bar{q}_0$ grows rapidly and (according to the numerical evidence) diverges as we approach the critical value of the ratio $N/T$, see Fig.\ \ref{figq0ES1}.

\begin{figure}
\begin{center}
\includegraphics[scale=0.55]{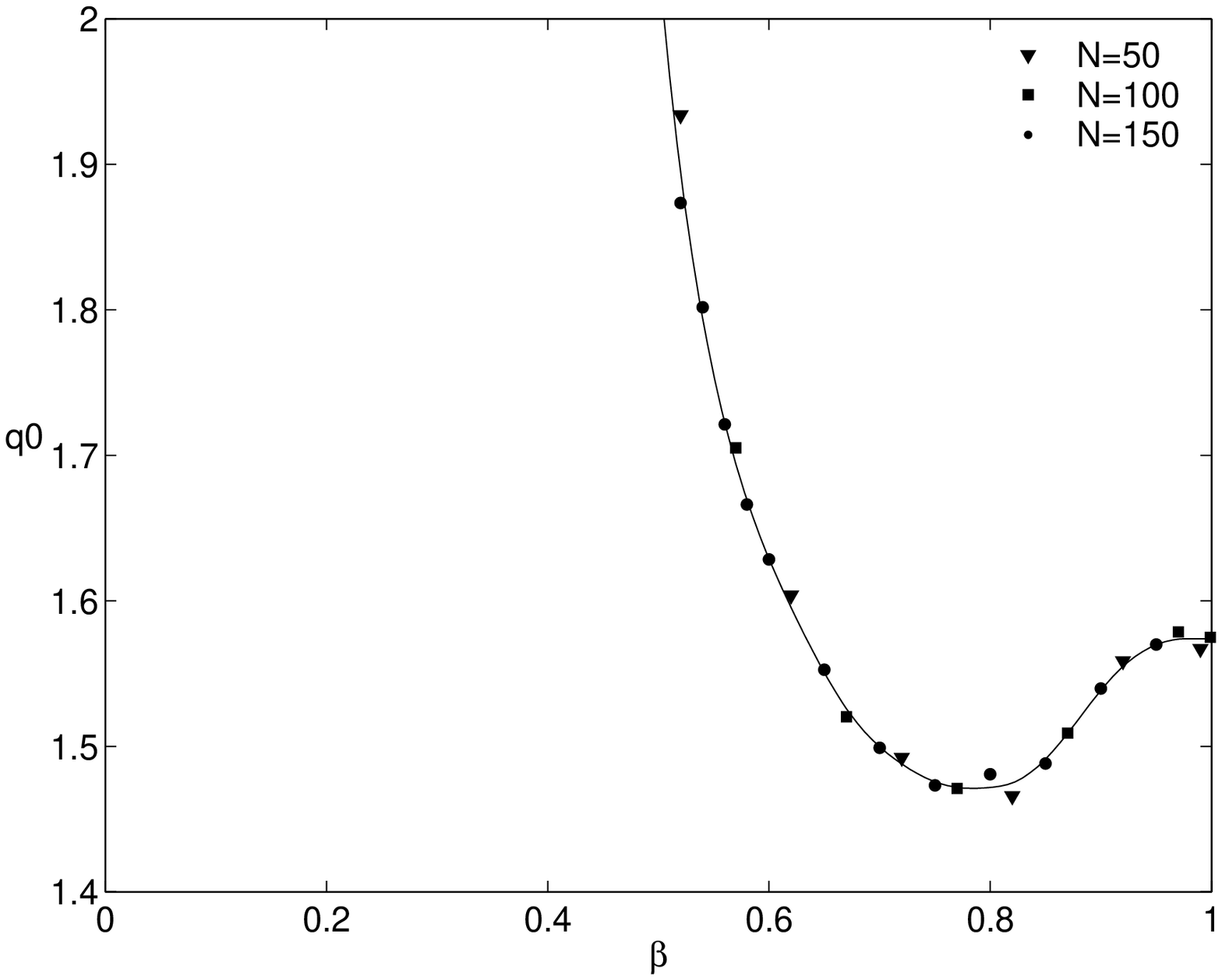}
\end{center}
\caption{$\bar{q}_{0,ES}$ as a function of $\beta$ for fixed $N/T=1/4$ (for various values of $N$).
\label{figq0ES2}}
\end{figure}

\begin{figure}
\begin{center}
\includegraphics[scale=0.55]{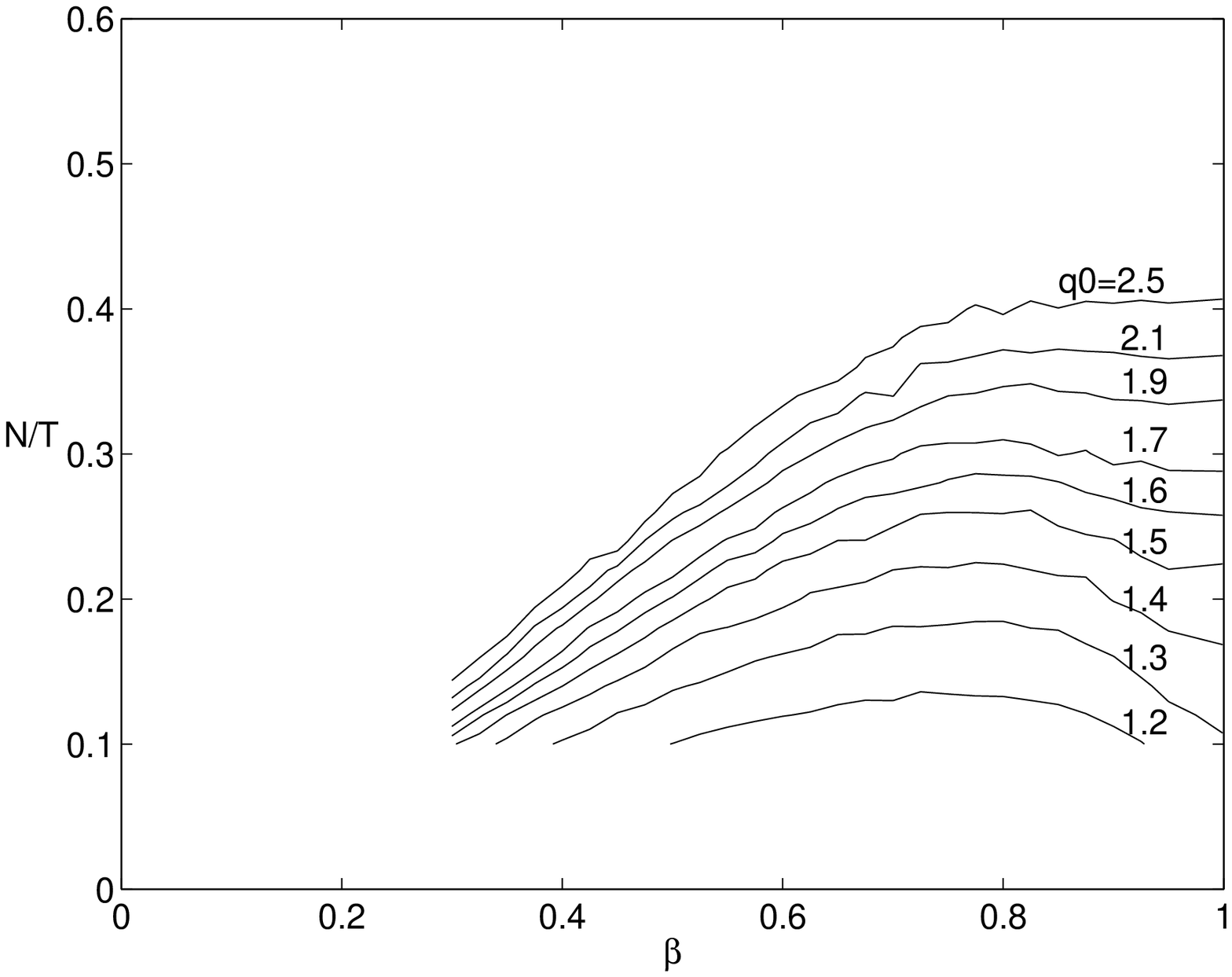}
\end{center}
\caption{Contour lines of $\bar{q}_0$ as a function of $\beta$ and $N/T$.
\label{figq0contour}}
\end{figure}

However, we encounter another surprising behaviour if we follow the variations of $q_0$ with $\beta$ for a fixed (sufficiently small) $N/T$, that is along a horizontal line in Fig.\ \ref{figpESphase}. As $\beta$ starts to decrease, first $\bar{q}_0$ decreases with it, but then it goes through a minimum and starts to grow until it diverges at the phase boundary, see Fig.\ \ref{figq0ES2}. This non-monotonic feature of $\bar{q}_0$ as a function of $\beta$ can be observed also in Fig.\ \ref{figq0contour}, where the level lines of the $\bar{q}_0(\beta,N/T)$ function have been represented.

\begin{figure}
\begin{center}
\includegraphics[scale=0.55]{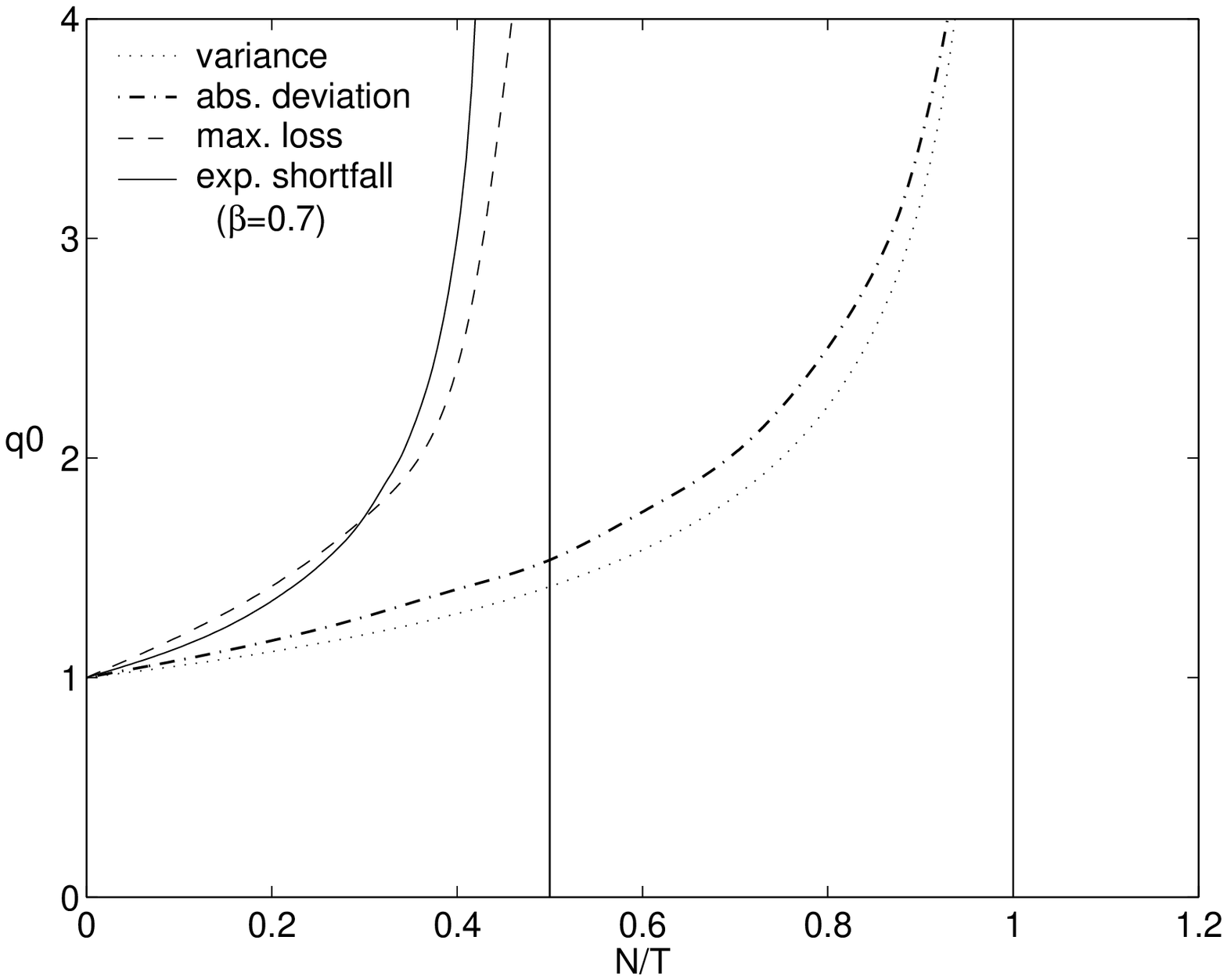}
\end{center}
\caption{$\bar{q}_0$ as a function of $N/T$ for all four risk measures discussed in the paper.
\label{figq0all}}
\end{figure}

We find the behaviour of both the phase boundary and $\bar{q}_0$ quite counterintuitive: we anticipated that expected shortfall would become less and less sensitive to noise as the confidence level (or threshold) $\beta$ decreases which means that we take into account more and more data. When it comes to a comparison between the sensitivities of the various risk measures, it seemed clear (we are obliged to C. Acerbi for a discussion on this point) that it is unfair to compair a high-$\beta$ ES to variance, because a high-$\beta$ ES sacrifices most of the observed data. What is unexpected is that with $\beta$ decreasing, the probability of a solution declines and that after an initial decrease $\bar{q}_0$ starts to grow.

In view of this nonmonotonic behaviour of $\bar{q}_0$, the most favourable comparison for ES would be to trace out the line on the $\beta$--$N/T$ plane along  which $\bar{q}_0$ is smallest. Unfortunately, this would require a huge computational effort that we have not been able to afford so far. On the basis of available evidence we strongly suspect, however, that ES would display an enhanced sensitivity to noise compared not only to variance but also MAD even under the most favourable circumstances. 

All the results in this section have been obtained by considering a portfolio of independent normal variables. It is easy to see, however, that they remain valid also for correlated normal random variables. A comparison of the sensitivities of all four risk measures (variance, absolute deviation, maximal loss, and expected shortfall) is shown in Fig.\ \ref{figq0all}.

\begin{figure}[h!]
\begin{center}
\includegraphics[scale=0.65]{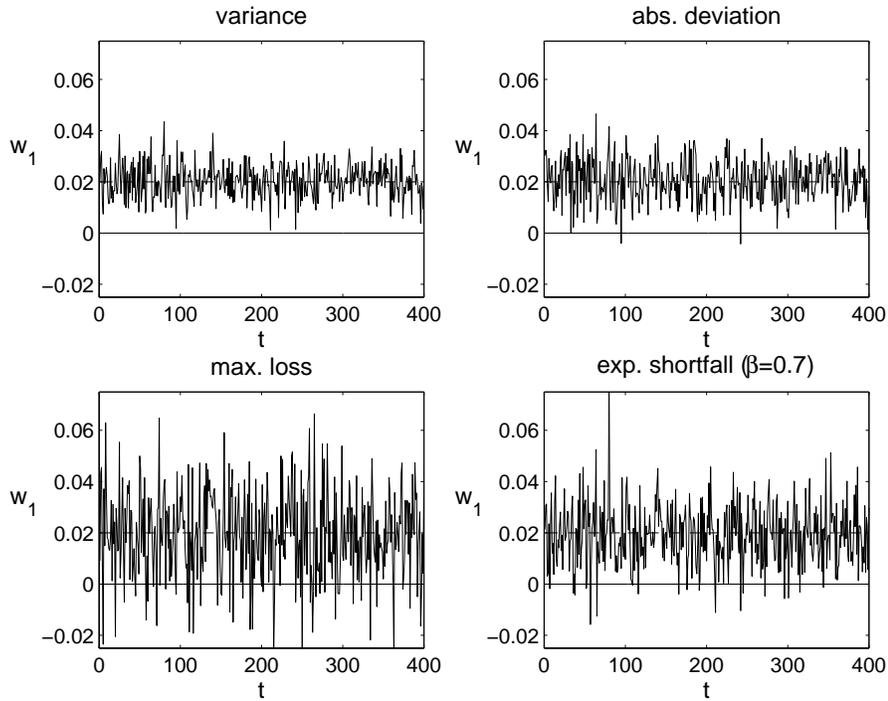}
\end{center}
\caption{$w_1^{*}$ as a function of time for all four risk measures, when time windows are non-overlapping ($N=50$, $T=500$).
\label{figw1newall}}
\end{figure}

\begin{figure}[h!]
\begin{center}
\includegraphics[scale=0.65]{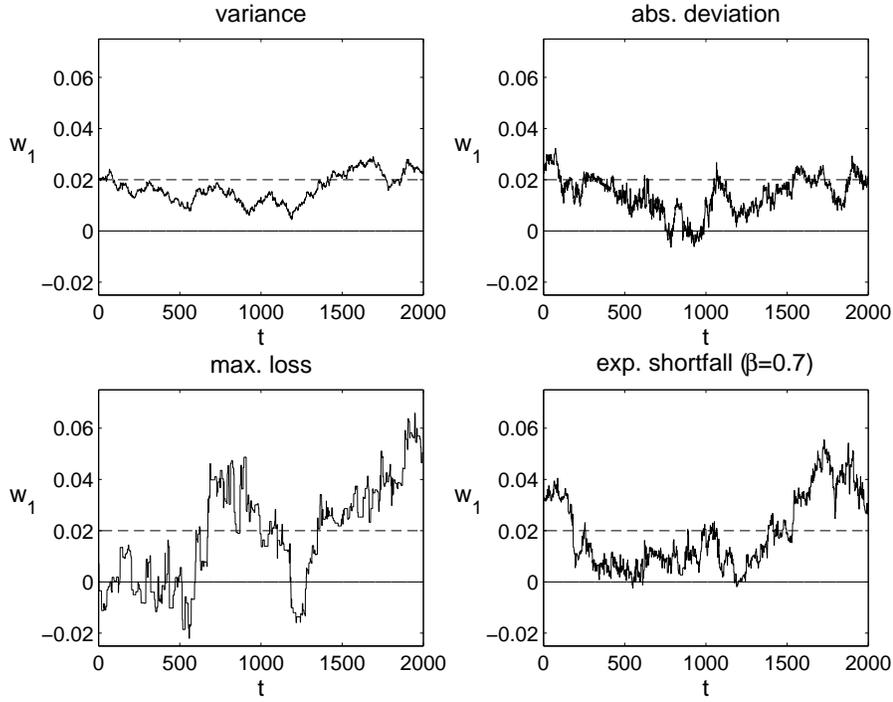}
\end{center}
\caption{$w_1^{*}$ as a function of time for all four risk measures, when the time window steps forward one unit at a time ($N=50$, $T=500$).
\label{figw1shiftall}}
\end{figure}

\begin{figure}
\begin{center}
\includegraphics[scale=0.55]{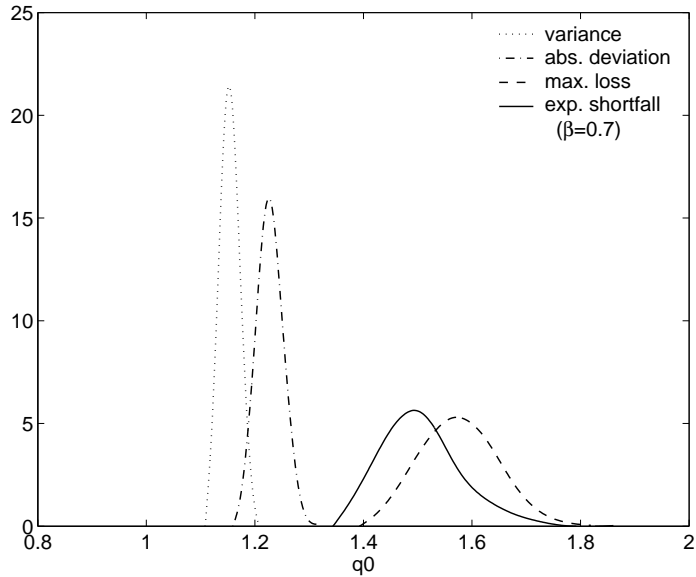}
\end{center}
\caption{Distribution of $q_0$ for $N=200$, $N/T=1/4$ for all the four risk measures discussed in the paper (simulation results).
\label{figq0histall}}
\end{figure}

\subsection{ Fluctuating weights and the distribution of $q_0$}

We can take a closer look at the noise-induced instability of portfolios again, and study the instability of weights and the probability distribution of $q_0$ as these quantities fluctuate from sample to sample. Repeating the same experiments (with exactly the same uncorrelated normal random variables for all four risk measures), we find the results depicted in Figs.\ \ref{figw1newall}--\ref{figw1shiftall}.
As we have seen already with the variance, the portfolio weights are quite unstable even when $\bar{q}_0$ is still acceptable. Sample to sample fluctuations with nonoverlapping windows are huge and uncorrelated, while those for windows sliding forward one step at a time show strong autocorrelations and tell a fictitious story of wandering weights whereas the true weights should stay constant. It is also clear that the instability of the weights becomes stronger and stronger as we go from one risk measure to the other, in the order: variance, MAD, ES and ML. 

The same conclusion is born out by the histograms of $q_0$ under various risk measures displayed in Fig.\ \ref{figq0histall}. Keeping everything else the same, we find that the distribution becomes wider and wider as we move from variance to maximal loss.


\section{Summary}

Due to the large number of assets in typical bank portfolios and the limited amount of data, noise is an all pervasive difficulty in portfolio theory. As such, it has attracted a lot of attention over the past decades. Most of these studies have focused on the case of variance as a risk measure, and on real life, empirical market data. In this paper we have studied and compared the noise sensitivity of portfolios optimized under various risk measures, including the variance but also absolute deviation, expected shortfall and maximal loss. We have approached the problem by introducing a global measure of portfolio sensitivity and testing the various risk measures by considering artificial, simulated portfolios of varying sizes $N$ and for different lengths $T$ of time series. 

Our findings demonstrate that the effect of noise is significant in all the investigated cases, it strongly depends on the ratio $N/T$, and actually diverges at a critical value of $N/T$ that depends on the risk measure in question. We have determined this critical value for all the four risk measures considered in this study and found that for the variance and MAD it is $N/T=1$, for ML it is $\frac{1}{2}$, and for ES it is smaller than $\frac{1}{2}$, its precise value depending on the threshold $\beta$, thereby forming a kind of ''phase boundary'', separating the region where, in the limit $N,T\to\infty$, optimization of the portfolio can be performed with probability one, from that where the probability of finding a solution is zero. We have also studied the smooth transition across this boundary in the case of expected shortfall and maximal loss for finite values of $N$ and $T$. Furthermore, we have measured the mean relative estimation error $\bar{q}_0$ for all four risk measures, and found that it diverges with a critical exponent -1/2 as we approach the critical point.

Another, more ''microscopic'' aspect of noise sensitivity is the sample to sample fluctuation of portfolio weights. We found large fluctuations already for portfolios optimized under variance, but the fluctuations for the other risk measures are stronger still. The instability of weights is conspicuous if we use non-overlapping samples; whereas overlapping samples generate autocorrelations and apparent structure in data that are uncorrelated by construction. We may stop for a moment at this point to muse over whether some of the apparent regularities we may discern in observing complex systems are not merely mirages generated by the lack of sufficient information.    

It clearly transpires from the present study that the other three risk measures pose a higher demand for input information than variance: for the same portfolio size and the same (normal) distribution of returns they require more data, i.e.\ longer time series, for them to be competitive. It is evident that for normally distributed returns and under the sensitivity measure $q_0$, variance is the best risk measure from the point of view of noise tolerance, which is, in a sense, a foregone conclusion. As empirical data are neither normal nor stationary, this ranking may, of course, change according to the given situation, but we believe that the effect of estimation error on real life portfolio selection can only be worse than in our artificial world.

Whereas in the case of variance a number of efficient filtering methods have been developed to mitigate the effect of noise, for alternative risk measures there are hardly any filtering methods available. In view of the enhanced noise sensitivity of portfolios optimized under these other risk measures, there is an obvious need for filtering methods adapted to the specific risk measures in question. 

In order to be able to formulate the main message of the paper, we had to make serious simplifying assumptions. One of them was the use of Gaussian-distributed return data.
It seems obvious to us that the existence of the phase transition, which is ultimately due to information deficit, does not depend on the particular distribution of returns, but it remains to be seen whether fat tailed (and/or correlated, and/or nonstationary) distributions will change the universality class of the transition.

Another simplifying feature was the omission of any constraints other than the budget constraint. It will be interesting to see how the re-introduction of the constraint on the expected return will modify the conclusions, especially as the estimation error of the returns may be far more serious than that of the covariances.

Perhaps the most unrealistic feature in the above treatment was the lack of any constraints on the portfolio weights. This goes to the heart of the matter: a ban on short selling or any other constraint that would make the domain of the optimization problem finite would evidently prevent the estimation error from blowing up, just as a finite volume eliminates a real phase transition in physical systems. As we argued at the end of Section \ref{chapvariance}, however, finite volume constraints would not resolve the problem of estimation error. Instead of running away to infinity along the unstable directions, the weights would stick to the "walls" of the allowed domain, in addition, they would jump around from wall to wall according to the random sample. This kind of behaviour can not provide a solid basis for rational decision making. 

We intend to return to all three of the omitted points mentioned above in a subsequent work. 

\ack

This paper is based on the lecture given by one of us (I.K.) at the 2005 Rome Summer School on ''Risk Measurement and Control''. He is obliged to the organizers of the School, Profs.\ G.\ Barone-Adesi, R.L.\ D'Ecclesia, and G.\ Szeg{\"o}, for the invitation and for prompting us to write up this paper. Most of the composition was done at Notre Dame University, Indiana, USA, where the authors enjoyed the hospitality of Prof.\ B.\ Jank{\'o} and the Institute of Theoretical Sciences. Valuable discussions with C.\ Acerbi and M. \ Mezard are also gratefully acknowledged. The authors have been partially supported by the ''Cooperative Center for Communication Networks Data Analysis'', a NAP project sponsored by the National Office of Research and Technology under grant No.\ KCKHA005.


\bibliography{noisy_risk_measures}
\bibliographystyle{elsart-harv}


\end{document}